\documentclass[12pt]{article}
\usepackage{amsmath,amssymb}
\usepackage{graphics}
\usepackage[dvipdfm]{graphicx}
\usepackage{color}

\makeatletter

\@addtoreset{equation}{section}
\makeatother

\begin{document}
\baselineskip=16pt
\begin{titlepage}
\begin{flushright}
%{\small ??}\\[-1mm]
{\small EPHOU-12-001}\\
{\small OU-HET 744/2012}\\
{\small TU-904}\\
%arXiv:1005.xxxx
\end{flushright}
\vspace*{1.2cm}

\begin{center}

{\Large\bf 
%Neutrinophilic Higgs in SUSY GUT
Phenomenology of SUSY $SU(5)$ GUT with neutrinophilic Higgs
} 
\lineskip .75em
\vskip 1.5cm

\normalsize
{\large Naoyuki Haba}$^1$,
{\large Kunio Kaneta}$^2$,   
and
{\large Yasuhiro Shimizu}$^3$ 

\vspace{1cm}

$^1${\it Department of Physics, 
Faculty of Science, Hokkaido University, Sapporo 060-0810, Japan}\\
$^2${\it Department of Physics, 
 Osaka University, Toyonaka, Osaka 560-0043, 
 Japan} \\
$^3${\it Department of Physics,
Tohoku University, Sendai 980-8578, Japan} \\

\vspace*{10mm}

{\bf Abstract}\\[5mm]
{\parbox{13cm}{\hspace{5mm}
%
%%%%%%%%%%%%%%%%%%%%%%%%%%%%%%%%%%%%%%%%%%%%%%%%%%%%%%%%%%%%%%%%
%             ABSTRACT                                         %
%%%%%%%%%%%%%%%%%%%%%%%%%%%%%%%%%%%%%%%%%%%%%%%%%%%%%%%%%%%%%%%%

Among three typical energy scales, 
 a neutrino mass scale ($m_\nu\sim$ 0.1 eV), 
 a GUT scale ($M_{GUT}\sim 10^{16}$ GeV), 
 and 
 a TeV-scale ($M_{NP}\sim 1$ TeV), 
 there is a fascinating relation of     
 $M_{NP}\simeq \sqrt{m_\nu\cdot M_{GUT}}$.  
The TeV-scale, $M_{NP}$, is a new physics
 scale beyond the standard model 
 which is regarded as 
 ``supersymmetry'' (SUSY) in this letter. 
%We suggest 
% a simple supersymmetric neutrinophilic 
% Higgs doublet model, 
We investigate phenomenology of SUSY $SU(5)$ GUT with neutrinophilic Higgs,
 which realizes the above relation dynamically as well as  
 the suitable %Dirac neutrino mass 
 magnitude of Dirac mass, $m_\nu$, 
 through  
 a tiny vacuum expectation value 
 of neutrinophilic Higgs.  
% without 
% additional scales other than 
% $M_{NP}$ and $M_{GUT}$.     
%A gauge coupling unification, which is an excellent feature
% in the supersymmetric standard model, 
% is preserved 
% automatically 
% in this setup. 
As a remarkable feature of this model,
 accurate 
 gauge coupling unification can be achieved 
 as keeping with a proton stability.
We also evaluate flavor changing processes
 in quark/lepton sectors.

}}

\end{center}

\vspace{.5cm}
\hspace*{1cm}
{\small 
PACS:
% 12.60.-i, 12.10.-g, 12.60.Fr
%12.10.-g, 12.10.Kt, 12.15.Ff
12.60.-i, 12.10.-g, 12.15.Ff
}

\end{titlepage}

%%%%%%%%%%%%%%%%%%%%%%%%%%%%%%%%%%%%%%%%%%%%%%%%%%%%%%%%%%%%
%                         Introduction                     %
%%%%%%%%%%%%%%%%%%%%%%%%%%%%%%%%%%%%%%%%%%%%%%%%%%%%%%%%%%%%
\section{Introduction}

There are 
 three typical energy scales, 
 a neutrino mass scale ($m_\nu\sim$ 0.1 eV), 
 a GUT scale ($M_{GUT}\sim 10^{16}$ GeV), 
 and  
 a TeV-scale ($M_{NP}\sim 1$ TeV)  
 which is a new physics 
 scale beyond the standard model (SM) and 
 regarded as 
 supersymmetry (SUSY) in this letter. 
Among these three scales, 
 we notice a fascinating relation,    
\begin{equation}
% M_{GUT}\cdot m_\nu \simeq M_{SUSY}^2
 M_{NP}^2 \simeq {m_\nu\cdot M_{GUT}}\ .  
\label{1}
\end{equation}
Is this relation an accident, or 
 providing a clue to the underlying 
 new physics ?
We take a positive stance toward 
 the latter possibility. 

As for a neutrino mass $m_\nu$, 
%The recent neutrino oscillation experiments 
% gradually reveal a 
% structure of 
% lepton sector\cite{Strumia:2006db, analyses}.  
%However, from the theoretical point of view, 
 its smallness is still a mystery, and 
 it is one of the
 most important clues to find new physics. % beyond the SM.
% standard model (SM). 
%A lot of ideas have been suggested to explain 
% the smallness of neutrino masses comparing to those of quarks and
% charged leptons. 
%How about considering a possibility that the 
Among a lot of possibilities, 
 a neutrinophilic Higgs doublet model 
 suggests an interesting explanation of 
 the smallness %of the neutrino masses 
 by %is originating from an extra Higgs
% doublet with
 a tiny vacuum expectation value (VEV) \cite{Ma}-\cite{Haba:2011fn}. 
This VEV from 
 a neutrinophilic Higgs doublet 
 is of ${\mathcal O}(0.1)$ eV 
 which is the same as the neutrino mass,
 so that it 
 suggests Dirac 
 neutrino\cite{Nandi,WWY,Davidson:2009ha,Davidson:2010sf}.\footnote{
In Refs.\cite{Ma,MaRa,Ma:2006km,HabaHirotsu,HabaTsumura,HS1,HS2}, 
 Majorana neutrino scenario is considered 
 through TeV-scale seesaw with  
 a neutrinophilic Higgs VEV of ${\mathcal O}(1)$ MeV.}  
Thus, the neutrino mass is much smaller than other fermions, 
 since its origin is 
 the tiny VEV from the different
 (neutrinophilic) Higgs doublet. 
Introduction of $Z_2$-symmetry 
 distinguishes the neutrinophilic Higgs from 
 the SM-like Higgs, 
 where $m_\nu$ is surely generated only through 
 the VEV of the neutrinophilic Higgs. 
The SUSY extension of the neutrinophilic Higgs doublet model 
 is considered in Refs.\cite{{Marshall:2009bk}, HS1, HS2, {Haba:2011fn}}. 
Since the 
 neutrino Yukawa couplings are not necessarily tiny anymore,
 some related researches have been done, such as,  
 collider phenomenology\cite{Davidson:2010sf,HabaTsumura}, 
 low energy thermal leptogenesis\cite{HS1, HS2},  
 cosmological constraints\cite{Sher:2011mx}\footnote{
A setup in Refs.\cite{Sher:2011mx} is different from 
 usual neutrinophilic Higgs doublet models, since 
 it includes a light Higgs particle.  %of order eV, 
% which usual 
 }, and so on.
%\footnote{ 
%The stability of a hierarchy between 
% the neutrinophilic Higgs VEV and 
% electroweak scale against radiative corrections  
% has been also done in Refs.\cite{Morozumi}. 
% }

On the other hand, 
 SUSY  
 is the most promising candidate 
 of new physics beyond the SM  
 because of a excellent success of gauge coupling unification. 
Thus, the SUSY SM well fits the GUT scenario as 
 well as an existence of 
 a dark matter candidate.

There are some attempts that try to realize 
 the relation in 
 Eq.(\ref{1}).
One example is 
 to derive $m_\nu$ from a higher dimensional 
 operator in the SUSY
 framework\cite{ArkaniHamed:2000bq}. 
Another example is to take a setup of 
 matter localization\cite{Agashe:2008fe}  
 in a warped extra dimension\cite{RS}. 
These scenarios are interesting, but 
 a model in this letter is much simpler 
 and contains no additional scales other than 
 $M_{NP}$, $m_\nu$, and $M_{GUT}$.     
(For other related papers, see, for example, 
 \cite{{related},{related2}}.)

%In this letter, we suggest a simple neutrinophilic 
% Higgs model in the SUSY GUT framework. 
%This model simply realizes the above relation from a $F$-flatness
% condition as well as $m_\nu\sim$ 0.1 eV.  

In this paper, 
% we suggest 
% a simple SUSY neutrinophilic Higgs doublet model, which 
% dynamically realizes the relation 
% of Eq.(\ref{1}).  
%{\color{magenta}
we investigate phenomenology of
a SUSY 
%{\color{magenta}
$SU(5)$ GUT with neutrinophilic Higgs ($SU(5)_{H_\nu}$) model 
%}
proposed in Ref. \cite{Haba:2011pm}.
\footnote{
%The author would like to thank R. Kitano 
% for pointing out a paper\cite{Kitano:2002px}, which 
% suggested the similar model and also estimated 
% lepton flavor violating processes.   
%{\color{magenta}
A similar model was suggested in Ref. \cite{Kitano:2002px}, 
where lepton flavor violation was also roughly estimated.
%}
} 
%}
Usually,
 SUSY neutrinophilic Higgs doublet models have %, 
% our model has no additional 
 tiny mass scale of 
 soft $Z_2$-symmetry breaking  
 ($\rho,\rho'={\cal O}(10)$ eV   
  %for ${\cal O}(1)$ TeV $B$-terms in 
 in Refs.\cite{HS1,HS2,{Haba:2011fn}}). 
This {\it additional} tiny mass scale 
 plays a crucial role of 
 generating the tiny neutrino mass, %in a 
% usual neutrinophilic Higgs doublet model, 
 however, 
% we did not argue 
 its origin %of it %this scale
 is completely unknown (assumption). 
In other words, 
 the smallness of $m_\nu$ is just replaced by 
 that of $Z_2$-symmetry breaking mass parameters, 
 and this is not an essential explanation of 
 tiny $m_\nu$.   
This is a common serious problem exists in 
 neutrinophilic Higgs doublet models in general. 
% has not 
% been argued. %can not know why this scale 
% ($|m_3^2|={\cal O}(0.1)$ MeV$^2$) 
% $\rho,\rho'={\cal O}(10)$ eV 
% appear?
%On the other hand, 
% the present model only has 
% two scales of
% $M_{GUT}$ and $M_{NP}$, 
% GUT ($10^{16}$ GeV) 
% and new physics ($1$ TeV), 
% and does not requir 
% any additional scales %such as 
% $10$ eV. 
% and naturally realize 
% the suitable magnitude of $m_\nu$  
%${\cal O}(0.1)$ eV
% naturally realized by 
%
% through Eq.(\ref{1}). 
%As for the
%
%Differently from 
% usual neutrinophilic doublet models, 
% our model has no additional 
% tiny mass scale 
% (artificial soft $Z_2$-symmetry breaking scale
%  of $|m_3^2|={\cal O}(0.1)$ MeV$^2$
% \cite{Nandi,WWY,Davidson:2009ha,Davidson:2010sf}). 
%This additional tiny mass plays a crucial role of 
% generating tiny neutrino mass in a 
% usual neutrinophilic Higgs doublet model, 
% however, 
% we can not know why this scale 
% ($|m_3^2|={\cal O}(0.1)$ MeV$^2$) 
% $\rho,\rho'={\cal O}(10)$ eV 
% appear?
%On the other hand, 
%Our model solves 
%{\color{magenta}
Notice, this problem can be solved by Ref. \cite{Haba:2011pm}
, where
%} 
% only has  
 two scales of
 $M_{GUT}$ and $M_{NP}$  
 naturally induce  
 the suitable magnitude of $m_\nu$  
 through the relation of Eq.(\ref{1}),   
 and does not require 
 any additional scales.  
%%%%%
The model contains a pair of new neutrinophilic Higgs doublets
 with GUT-scale masses, and 
 the $Z_2$-symmetry is broken by TeV-scale 
 dimensionful couplings
 of these new doublets to the ordinary SUSY
 Higgs doublets. 
Once the ordinary Higgs doublets obtain VEVs $(v_{u,d})$ by the usual
 electroweak symmetry breaking,
 they trigger VEVs for the neutrinophilic 
 Higgs doublets of 
 $v_{u,d}M_{NP}/M_{GUT} (\sim m_\nu)$. 
Then, ${\cal O}(1)$ Yukawa
 couplings of the neutrinophilic doublets to
 $L\ N$
 ($L$: lepton doublet, $N$: right-handed neutrino)  
 give neutrino
 masses of the proper size.
%%%
%A gauge coupling unification %, 
% which is an excellent feature
% in the SUSY SM, 
% is also preserved automatically 
% in our 
% setup. 
We can also obtain
 a GUT embedding of the SUSY neutrinophilic Higgs doublet model,
 which realizes the relation, $m_\nu \sim v_{u,d}M_{NP}/M_{GUT}$, dynamically.
As a remarkable feature of this model,
 accurate gauge couplings can be unified as keeping a proton stability. 
% which is one of the main topics in this paper. 
Flavor changing processes are also sensible aspect of this model.
In general,
 flavor violation in charged lepton sector is related to that in quark sector 
 because lepton doublet and right-handed down-type quark
 are contained in a same multiplet in $SU(5)$ GUT.
Particularly,
 neutrino oscillation directly contributes flavor violations in both sectors.
It is one of our purposes to evaluate such flavor violating processes.

This paper is organized as follows.
In section 2,
we review a SUSY $SU(5)_{H_\nu}$ model.
In section 3 and 4, 
we discuss a gauge coupling unification, 
and investigate flavor violations in SUSY 
%{\color{magenta}
$SU(5)_{H_\nu}$ model.
%}
These sections are main parts of this paper.
In section 5,
 we present a summary.

\section{SUSY $SU(5)$ GUT with neutrinophilic Higgs}

Before showing 
 a SUSY
% {\color{magenta}
  $SU(5)_{H_\nu}$ model 
%  }
  \cite{Haba:2011pm}, 
 we show a SUSY neutrinophilic Higgs doublet 
 model at first.  
This has 
 a specific parameter region which is different from
 Refs.\cite{HS1, HS2, {Haba:2011fn}}. 
We introduce $Z_2$-parity, where 
 only vector-like neutrinophilic Higgs doublets and 
 right-handed neutrino have odd-charge. 
The superpotential of the Higgs sector 
 is given by
\begin{eqnarray}
\mathcal{W}_h = 
 \mu H_u H_d  + M H_{\nu} H_{\nu'} - \rho H_u H_{\nu'} - \rho' H_{\nu} H_d.
\label{WW}
\end{eqnarray}
%where 
$H_\nu$ ($H_{\nu'}$) is a neutrinophilic 
 Higgs doublet, % superfield,  
 and $H_\nu$ has Yukawa interaction 
 of $LH_\nu N$, which induces a tiny Dirac neutrino 
 mass through the tiny VEV of   
 $\langle H_\nu\rangle$.  
This is the origin of smallness of 
 neutrino mass,  
 and this paper devotes a Dirac neutrino scenario, i.e., 
% where neutrino mass is given by 
 $m_\nu \simeq \langle H_\nu\rangle ={\cal O}(0.1)$ eV.  
% whose 
% magnitude is of 
On the other hand, 
 $H_{\nu'}$ does not couple with any matters. 
$H_u$ and $H_d$ are 
 Higgs doublets in 
 the minimal SUSY SM 
 (MSSM), and  
 quarks and charged lepton obtain their masses 
 through 
 $\langle H_u\rangle$ and $\langle H_d\rangle$. 
Note that this structure is guaranteed by 
 the $Z_2$-symmetry.  
%In this paper, 
Differently from conventional 
 neutrinophilic Higgs doublet models, 
 we here take $M$ as the GUT scale  
 and $\mu, \rho, \rho'$  %other mass parameters as  
 ${\cal O} (1)$ TeV. 
%We assume that
The soft $Z_2$-parity breaking parameters, 
 $\rho$ and $\rho'$, might be  
 % terms, $\rho H_u H_{\nu'}$ and $\rho' H_{\nu} H_d$, are 
% expected to be 
 induced from SUSY breaking effects  
 %(which will be discussed in the next section),
 ({\it see below}),
 and %here  
 we %just
 regard   
 $\rho$ and $\rho'$ as 
% just 
 mass 
 parameters of  
 new physics scale, $M_{NP}={\cal O}(1)$ TeV. 
Remind that usual 
% Notice that these magnitudes are completely different from
 SUSY neutrinophilic doublet models take %has a setup %use %, where 
% $M={\cal O}(1)$ TeV and  
 $\rho,\rho'={\cal O}(10)$ eV 
 (for ${\cal O}(1)$ TeV $B$-terms)\cite{HS1, HS2, {Haba:2011fn}}. 
This additional tiny mass scale plays a crucial role of 
 generating the tiny neutrino mass  
 however, 
 its origin 
 is just an assumption. 
Thus, 
 the smallness of $m_\nu$ is just replaced by 
 that of $\rho$ and $\rho'$.  
This is a common serious problem exists in 
 neutrinophilic Higgs doublet models in general. 
The present model solves this problem, in which 
 two scales of
 $M_{GUT}$ and $M_{NP}$  
 induce  
 the suitable magnitude of $m_\nu$
 dynamically,  
% through the relation of Eq.(\ref{1}),   
 and does not require 
 any additional scales, such as 
 ${\cal O}(10)$ eV. 
% any additional scales. 
%On the other hand, 
% the present model only has 
% two scales of GUT (${\cal O}(10^{16})$ GeV) 
% and new physics (${\cal O}(1)$ TeV), 
% and does not require 
This is one of the excellent points 
% in our model.  
%{\color{magenta}
in this model.
%}

Amazingly, 
 stationary conditions make 
 the VEVs of % $H_\nu$ and $H_{\nu'}$ become 
 neutrinophilic Higgs fields be  
\begin{eqnarray}
v_{\nu} =  \frac{\rho  v_u}{M} , \;\;\;
v_{\nu'} = \frac{\rho' v_d}{M} .  
\label{wvev}
\end{eqnarray}
It is worth noting that 
 they are induced dynamically through the 
 stationary conditions, % in Eqs.(\ref{ssc3}) and (\ref{ssc4}),   
%which 
 and their magnitudes are surely 
 of ${\cal O}(0.1)$ eV.  
% since $M={\cal O}(10^{16})$ GeV,
% $\rho, \rho'$ are of order TeV scale, 
% and $v_{u,d}$ are the weak scale.    
%
Since the masses of neutrinophilic Higgs $H_\nu$ and 
 ${H}_{\nu'}$ are super-heavy as the GUT scale, 
 there are no other vacua (such as,  
 $v_{u,d} \sim v_{\nu, \nu'}$)
 except for 
 $v_{u,d} \gg v_{\nu, \nu'}$ \cite{Haba:2011fn}. 
Also, 
 their heaviness guarantees the stability 
 of the VEV hierarchy, 
 $v_{u,d} \gg v_{\nu, \nu'}$, against radiative 
 corrections \cite{Morozumi, Haba:2011fn}. 
It is because, 
 in the effective potential,  
 $H_\nu$ and 
 ${H}_{\nu'}$ inside 
 loop-diagrams are suppressed 
 by their GUT scale masses. 
%Anyhow, 
% we stress again that 
% the relation of Eq.(\ref{1}) is 
% dynamically obtained in 
%% of $M_{GUT}\cdot m_\nu \simeq M_{SUSY}^2$ in 
% Eq.(\ref{rel2}). 

The model 
%we suggested 
%{\color{magenta}
suggested in Ref. \cite{Haba:2011pm}
%} 
has 
 the GUT scale mass of neutrinophilic Higgs doublets 
% are of the GUT scale
 in Eq.(\ref{WW}), 
 so that it is naturally embeded into 
% shown in the previous section
 a GUT framework, and  
 it is the SUSY $SU(5)_{H_\nu}$ model. 
A superpotential of a Higgs sector 
 at the GUT scale  
 is given by
\begin{eqnarray}
\mathcal{W}_H^{\rm GUT} = 
 M_0 {\rm tr}\Sigma^2 + \lambda {\rm tr}\Sigma^3 + 
 H \Sigma \bar{H} + \Phi_\nu \Sigma \bar{\Phi}_\nu
 - M_1 H \bar{H}  - M_2 \Phi_\nu \bar{\Phi}_\nu  ,
\label{WGUT}
\end{eqnarray}
where $\Sigma$ is an adjoint Higgs whose VEV reduces the 
 GUT gauge symmetry into the SM. 
%$M_{0,1,2}$ are the GUT scale mass parameters.  
$\Phi_\nu$ ($\bar{\Phi}_\nu$) is a neutrinophilic 
 Higgs of (anti-)fundamental 
 representation, which contains $H_\nu$ ($H'_\nu$) 
 in the doublet component    
 (while the triplet component is denoted as $T_\nu$ ($\bar{T}_\nu$)). 
$\Phi_\nu$ and $\bar{\Phi}_\nu$ are odd under 
 the 
 $Z_2$-parity. 
$H$ ($\bar{H}$) is a 
 Higgs of (anti-)fundamental 
 representation, which contains $H_u$ ($H_d$) 
 in the doublet component 
 (while the triplet component is denoted as $T$ ($\bar{T}$)). 
The VEV of $\Sigma$ and $M_{0,1,2}$ are all  
 of ${\cal O}(10^{16})$ GeV, 
 thus we encounter so-called triplet-doublet (TD)
 splitting problem. 
Some mechanisms have been suggested 
 for a solution of TD splitting, 
 but here 
 we show a case that 
%  Here we do not take a explicit 
% mechanism for the solution of TD 
% splitting problem, and 
% assume its sucess and consider 
% the effective superpotential 
% below the GUT scale. 
%The simplest example of
 the TD splitting is realized 
 just by a fine-tuning 
 between $\langle \Sigma \rangle$ and $M_1$. 
That is, 
 $\langle \Sigma \rangle - M_1$ induces 
 GUT scale masses of $T, \bar{T}$, 
 while 
 weak scale masses of $H_u, H_d$. 
This is a serious fine-tuning, 
 so that we can not expect a simultaneous 
 fine-tuned cancellation %also happens 
 between 
 $\langle \Sigma \rangle$ and $M_2$.  
Thus, we %naturally 
 consider a case that the TD splitting only %fine-tuning only   
 works in %happens between
 $H$ and $\bar{H}$, 
 while not works in %happen between %neutrinophilic Higgs fields, 
 $\Phi_\nu$ and $\bar{\Phi}_\nu$. 
This situation makes Eq.(\ref{WGUT}) become 
% where the TD splitting is finished 
% in the MSSM Higgs fields, 
% an effective superpotential 
% is given by 
%
\begin{eqnarray}
\mathcal{W}^{eff}_H = 
 \mu H_u H_d 
 + M H_{\nu} H_{\nu'} 
 + M'T \bar{T}
 + M'' T_\nu \bar{T}_\nu.
\label{WGUT2}
\end{eqnarray}
This is the effective superpotential of the Higgs sector  
 below the GUT scale, and   
 $M, M', M''$ are of ${\cal O}(10^{16})$ GeV, while %GUT scale masses, while  
 $\mu={\cal O} (1)$ TeV.

%We assume triplet components do not take VEVs 
% so that we omit them, for simplicity. 
%Actually, if similar terms of 
% $- \rho T_u T_{\nu'} - \rho' T_{\nu} T_d$ 
% are introduced in the triplets 
% ($T$s are triplets), 
% they obtain tiny VEVs as doublets.
%However, this is the color and electro-magnetic 
% symmetry breaking, so that 
% we assume there are no above terms 
% for triplets. 

Now let us consider 
 an origin of  
 soft $Z_2$-parity breaking 
 terms, $\rho H_u H_{\nu'}$ and $\rho' H_{\nu} H_d$ in 
 Eq.(\ref{WW}). 
They play a crucial role of 
 generating
 the marvelous relation in Eq.(\ref{1})
 as well as 
 a tiny Dirac neutrino mass. 
Since the values of %magnitudes of 
 $\rho, \rho'$ are 
 of order ${\cal O}(1)$ TeV, 
% which means 
 they might be induced from %its origin should be 
 the SUSY breaking effects. 
%For the origin, 
We can consider some 
 possibilities for this mechanism. %it. %this setup. %derivation. 
% as underlying theories. 
%
One example is to 
 take a non-canonical K$\ddot{\rm a}$hler of  
 $[S^\dagger (H_uH_{\nu'} + H_\nu H_d)+{\rm h.c.}]_D$. 
Where $F$-term of $S$ could induce 
 the $\rho$- and $\rho'$-terms effectively 
 through the SUSY breaking scale 
 as in 
 Giudice-Masiero mechanism\cite{GM}.   
%Actually, we should include 
% the $S$ for the following potential 
% analysis, however, 
% it depends on the model %mechanism to
% of realizing the $\rho$- and $\rho'$-terms from 
% $\langle S \rangle$, and 
% this is not a main topics 
% of this paper. 
%Also, 
There might be other models 
 which induce the $\rho$- and $\rho'$-terms
 in Eq.(\ref{WW}) except for 
 introducing a singlet $S$.  
%Thus, we use  
% $\rho$ and $\rho'$ as 
% mass 
% parameters of  
% ${\cal O}(1)$ TeV in the following analyses.  
%Also we might consider 
% a framework of extradimension and 
% sequestering to explain 
% why other couplings with $S$ such as 
% $SH\bar{H}$ are negligible. 
%The light singlet sometimes causes
% hierarchy problems, and 
% we do not deal with the origin of 
% $\rho, \rho'$ here. 

%\section{Phenomenology}

\section{Gauge coupling unification and proton-decay}

%As for the
% gauge coupling unification, 
% it is preserved automatically 
% in our
% setup, since 
% fields except for 
% the MSSM have masses of 
% order the GUT scale.
In this section, we discuss
 a characteristic feature of 
 the gauge coupling unification
 and the proton-decay 
 in SUSY 
 %{\color{magenta}
 $SU(5)_{H_\nu}$ model
 %}
 by focusing on a role of $T_\nu$ and $\bar{T}_\nu$. 
As for 
 the minimal SUSY $SU(5)$ GUT model, 
 in order to unify the gauge couplings,
 mass of $T$ and $\bar{T}$ should be lighter than 
 the GUT scale as 
 $3.5\times 10^{14}\;{\rm GeV}\stackrel{<}{_\sim}
 M'\stackrel{<}{_\sim}3.6\times 10^{15}$ GeV 
 due to threshold corrections \cite{Murayama:2001ur}.
However, to avoid the rapid proton decay, 
  $M'$ must be heavier than the GUT scale ($M'>M_{GUT}$).  
Hence, it is difficult to achieve both 
 accurate gauge coupling unification and 
 enough proton stability in the 
 minimal SUSY $SU(5)$ GUT.

A situation becomes different 
 in the SUSY
 %{\color{magenta}
  $SU(5)_{H_\nu}$ model.  
 % }
%gauge couplings can be unified while a proton is stable.
%We devote this section to considering this property. 
In this model, 
 a superpotential of the Yukawa sector 
 is given by
\begin{align}
{\cal W}_Y&=\frac14 f_{u_{ij}} \psi_i \psi_j H  
  + \sqrt{2} f_{d_{ij}} \psi_i \phi_{j} \bar{H}  
 + f_{\nu_{ij}} \eta_i \phi_{j} \Phi_{\nu}\label{Wmatter}
\end{align}
 at the GUT scale, 
 where $i$ and $j$ are family indices.
$\psi_i,\; \phi_i,$ and $\eta_i$ are 10-plet, $\bar 5$-plet, and singlet
 in $SU(5)$ gauge group, respectively, 
 which are 
% The superpotential, Eq. (\ref{Wmatter}), can be
 written in terms of MSSM fields as
\begin{align}
&\psi_i = \{Q_i,e^{-i \phi_{u_i}}U_i, (V_{KM})_{ij}{\bar E}_j\},\;\;
\phi_i = \{(V_{D})_{ij} {\bar D}_j, (V_{D})_{ij}L_j\},\;\;
\eta_i = \{e^{-i \phi_{\nu_i}} {\bar N}_i\}.
\end{align}
Since Yukawa couplings are written as 
\begin{align}
&f_{u_{ij}}=f_{u_i}e^{i\phi_{u_i}} \delta_{ij},\;\;
f_{d_{ij}}=(V^*_{KM})_{ik} f_{d_k} (V^{\dagger}_D)_{kj},\;\;
f_{\nu_{ij}}=f_{\nu_i}e^{i\phi_{\nu_i}} \delta_{ij}, 
\end{align}
 the superpotential 
 in this basis 
 is given by
\begin{align}
{\cal W}_Y=
&f_{u_i} Q_i \bar{U}_i H_u
  + (V_{\rm KM}^{*})_{ij} f_{d_j} Q_i \bar{D}_{j} H_d+f_{d_i} \bar{E}_i L_i H_d\nonumber\\
&+ f_{u_j} (V_{\rm KM})_{ji} \bar{E}_i \bar{U}_j T -\frac12 f_{u_i}Q_i Q_i T
+ (V_{\rm KM}^{*})_{ij} f_{d_j}  
\bar{U}_i \bar{D}_j \bar{T}
 - (V_{\rm KM}^{*})_{ij} f_{d_j} Q_i L_j \bar{T}\nonumber\\
&- f_{\nu_i} (V_{D})_{ij} \bar{N}_i L_j H_{\nu}
+ f_{\nu_i} (V_{D})_{ij} \bar{N}_i \bar{D}_j T_{\nu}
,\label{Wsusygut}
\end{align}
where CP phases, $\phi_{u_i}$ and $\phi_{\nu_i}$, are omitted,
 for simplicity. 
The terms from the fourth to seventh in Eq. (\ref{Wsusygut})
 cause proton-decay, % processes.
%These contributions are the same as minimal SUSY $SU(5)$ GUT case.
%{\color{cyan} 
 which also exist in the minimal SUSY $SU(5)$ GUT. 
Thus, we should take $M'>M_{GUT}$ to avoid the rapid proton decay.
%}
Meanwhile the last term in Eq. (\ref{Wsusygut})
 has nothing to do with the proton decay. 
Since $T_\nu$ and $\bar{T}_\nu$ 
 contribute beta functions of
 $SU(3)_c \times U(1)_Y$,  %the strong coupling $\alpha_s$ and .
 accurate gauge coupling unification 
 is achieved with $T_\nu$ and $\bar{T}_\nu$ threshold corrections  
 with 
 $3.5\times 10^{14}\;{\rm GeV}\stackrel{<}{_\sim} M'' 
 \stackrel{<}{_\sim}3.6\times 10^{15}$ GeV. 
%if we take $m_T\stackrel{>}{_\sim}$ and $m_{T_\nu}\sim 5\times 10^{14}$ GeV,
Therefore, the 
 SUSY 
 %{\color{magenta}
 $SU(5)_{H_\nu}$ model
 %}
 can realize not only 
 the accurate gauge coupling unification but also 
 the proton stability.
Remembering that $M$ is the GUT scale,  
 ${\cal O}(1)$ \% tuning 
 between $M$ and $M''$ is needed, but  
 it can happen. 
Or, no tuning is required 
 when
 %{\color{magenta}
  one of couplings is
  %}
 of ${\cal O}(0.01)$,
 %{\color{magenta}
for example a coupoing of $S^\dagger H_u H_{\nu'}$.  
%}

\section{Flavor changing processes}

%Flavor violations in charged lepton sector and quark sector are related with each other
Flavor changing in the lepton sector is related to that in the quark
 sector, 
 since $L$ and $D$ 
 are contained in a same multiplet in $SU(5)_{H_\nu}$. 
Where, mixing angles in $V_D$ are expected to be large, 
 and 
 masses of left-handed slepton and right-handed down-type squark
 get sizable radiative corrections 
 in off-diagonal elements of flavor space. 
Leading $\log$ approximation makes 
 the off-diagonal elements %of the mass matrices are given by
\begin{align}
 (\delta m^2_{\tilde{L}})_{ij}&\simeq
-\frac{1}{8\pi^2}f^2_{\nu_k}(V^*_{D})_{ki}(V_{D})_{kj}(3m_0^2+A_0)
\log\frac{M_P}{M},\label{dmL}\\
 (\delta m^2_{\tilde{D}})_{ij}&\simeq
-\frac{1}{8\pi^2}f^2_{\nu_k}(V^*_{D})_{ki}(V_{D})_{kj}(3m_0^2+A_0)
\log\frac{M_P}{M''},\label{dmD}
\end{align}
where $M_P$ is the Planck scale, 
 $m_0$ and $A_0$ are universal scalar mass and 
 trilinear coupling in mSUGRA scenario.
%$\delta m_{\tilde{L}}^2$
Equation (\ref{dmL}) 
 originates from a loop diagram of $N$ and $H_\nu$, 
 where an energy scale in renormalization group equations %(RGEs)
 runs from $M_P$ to $M$ ($H_\nu$, $\bar{H}_\nu$
 mass). 
%because we assume that $H_\nu$ is at the GUT scale.
On the other hand, 
 Eq.(\ref{dmD}) is induced from 
 a loop diagram of $N$ and $T_\nu$, which runs from 
 $M_P$ to $M''$ ($T_\nu$, $\bar{T}_\nu$ 
 mass).  
% and $T_\nu$ is below the GUT scale.
Notice that loop effects of Eqs.(\ref{dmL}) and (\ref{dmD})
 are different from 
 those in $SU(5)$ with right-handed
 neutrinos ($SU(5)_{RN}$).\footnote{See, for example, \cite{Hisano:1998fj}}
In $SU(5)_{RN}$ model,
 neutrinos are Majorana and 
 counterparts of Eqs.(\ref{dmL}) and (\ref{dmD}) are given by
\begin{align}
 (\delta m^2_{\tilde{L}})_{ij}&\simeq 
-\frac{f_{\nu_k}f_{\nu_m}}{8\pi^2}
(V_D^*)_{ki}(V_M^*)_{lk}(V_M)_{lm}(V_D)_{mj}(3m_0^2+A_0)
\log\frac{M_P}{M_{N_l}},\label{dmL-RN}\\
 (\delta m^2_{\tilde{D}})_{ij}&\simeq
-\frac{1}{8\pi^2}f^2_{\nu_k}
(V_D^*)_{ki}(V_D)_{kj}(3m_0^2+A_0)
\log\frac{M_P}{M'},\label{dmD-RN}
\end{align}
%{\color{cyan}
where $M_{N_l}$ is a 
%{\color{magenta}
diagonal
%}
 Majorana mass of $N_l\; (l=1, 2, 3)$.
% an eignvalue (?) 
%a mass of
 %of a mass matrix of right-handed neutrinos 
 %which 
 A mass matrix of $N$ is diagonalized by a unitary
 matrix $V_M$ \cite{Hisano:2003bd}, which 
% }
 does not appear in Eq.(\ref{dmD-RN}) 
 because $M_{N_l}$ is usually assumed to be smaller than $M'$ ($> M_{GUT}$).
By comparing the $SU(5)_{RN}$ model  
 with the 
 %{\color{magenta}
 $SU(5)_{H_\nu}$ model, 
 %}
%we can find two remarkable points in SUSY neutrinophilic Higgs GUT.
 we can find an advantage point in the latter model.
 %SUSY neutrinophilic Higgs GUT.
%One is that predictability is improved in SUSY neutrinophilic Higgs GUT
This is 
 a predictability, that is,
 flavor changing processes are strongly predicted %more violations is improved
 since there are no degrees of freedom of $V_M$. 
This means that flavor violations in charged lepton sector
 are directly related to  
 those in the quark sector
 through 
 the large flavor mixings in the neutrino sector.
%We should also mention a remarkable point as follows.
And, 
 even if the mass matrix of the right-handed neutrinos is diagonal, 
 magnitudes of 
 $\delta m^2_{\tilde{L}}$ and $\delta m^2_{\tilde{D}}$ 
 in the 
% {\color{magenta}
 $SU(5)_{H_\nu}$ model
 %}
 are different from those in $SU(5)_{RN}$ model.  
%If mass matrix of right-handed neutrinos is diagonal, 
%$\delta m^2_{\tilde{L}}$ and 
%In this situation, 
%$\delta m^2_{\tilde{D}}$ in $SU(5)_{RN}$ are the same as 
%SUSY neutrinophilic Higgs GUT case excepting $\log$ factors.
%
%Excepting $\log$ factors,
%$\delta m^2_{\tilde{L}}$ and $\delta m^2_{\tilde{D}}$ in $SU(5)_{RN}$ are the same as 
%those in SUSY neutrinophilic Higgs GUT
%if mass matrix of right-handed neutrinos is diagonal.
For example, 
%$\delta m^2_{\tilde{D}}$ in $SU(5)_{RN}$ 
 Eq.(\ref{dmD-RN})
 can be a few percent smaller than
 that in 
% {\color{magenta}
 $SU(5)_{H_\nu}$ model
% }
 due to their $\log$ factors.
The magnitude of the $\log$ factor is 
 $\log\frac{M_P}{M'}\stackrel{<}{_\sim} \log\frac{M_P}{M''}$, 
 since $M'$ must be larger than the GUT scale for the proton stability 
 and $M''$ must be of ${\cal O}(10^{14})$ GeV for the accurate 
 gauge coupling unification.

\begin{figure}[htbp]
  \def\@captype{table}
  \begin{minipage}[r]{.48\textwidth}
  \includegraphics[width=80mm]{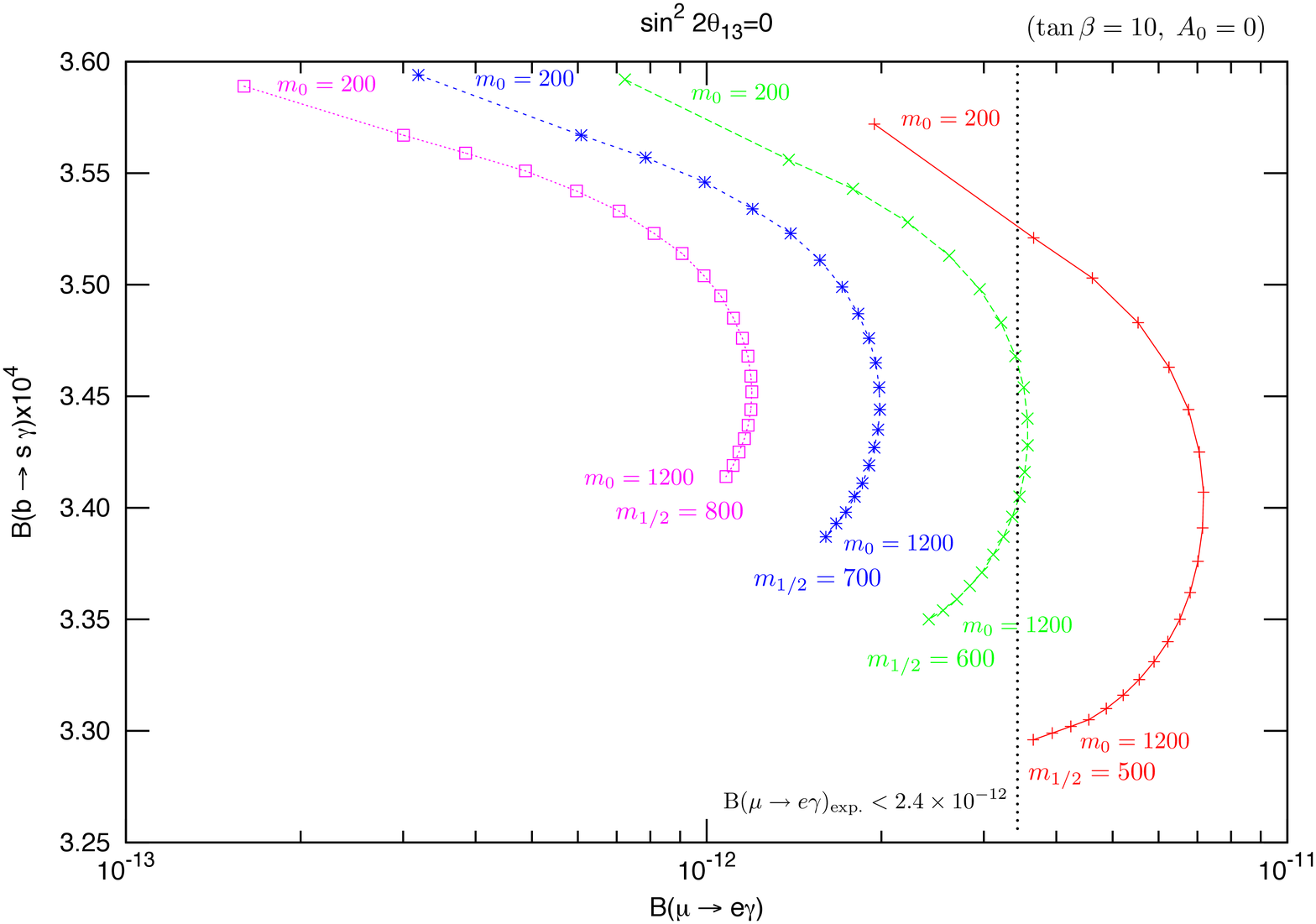}
\caption{\footnotesize Contour plot of ${\rm B}(b\to s\gamma)$ and ${\rm B}(\mu\to e\gamma)$ with $\sin^22\theta_{13}=0$. Here, we take $200{\rm GeV}\leq m_0\leq 1200{\rm GeV}$, $500{\rm GeV}\leq m_{1/2}\leq 800{\rm GeV}$, $A_0=0$, and $\tan\beta=10$. Experimental upper bound for ${\rm B}(\mu\to e\gamma)$ is $2.4 \times 10^{-12}$.}
 \label{meg-bsg-0}
\end{minipage}
  \hfill
  \begin{minipage}[l]{.48\textwidth}
  \includegraphics[width=80mm]{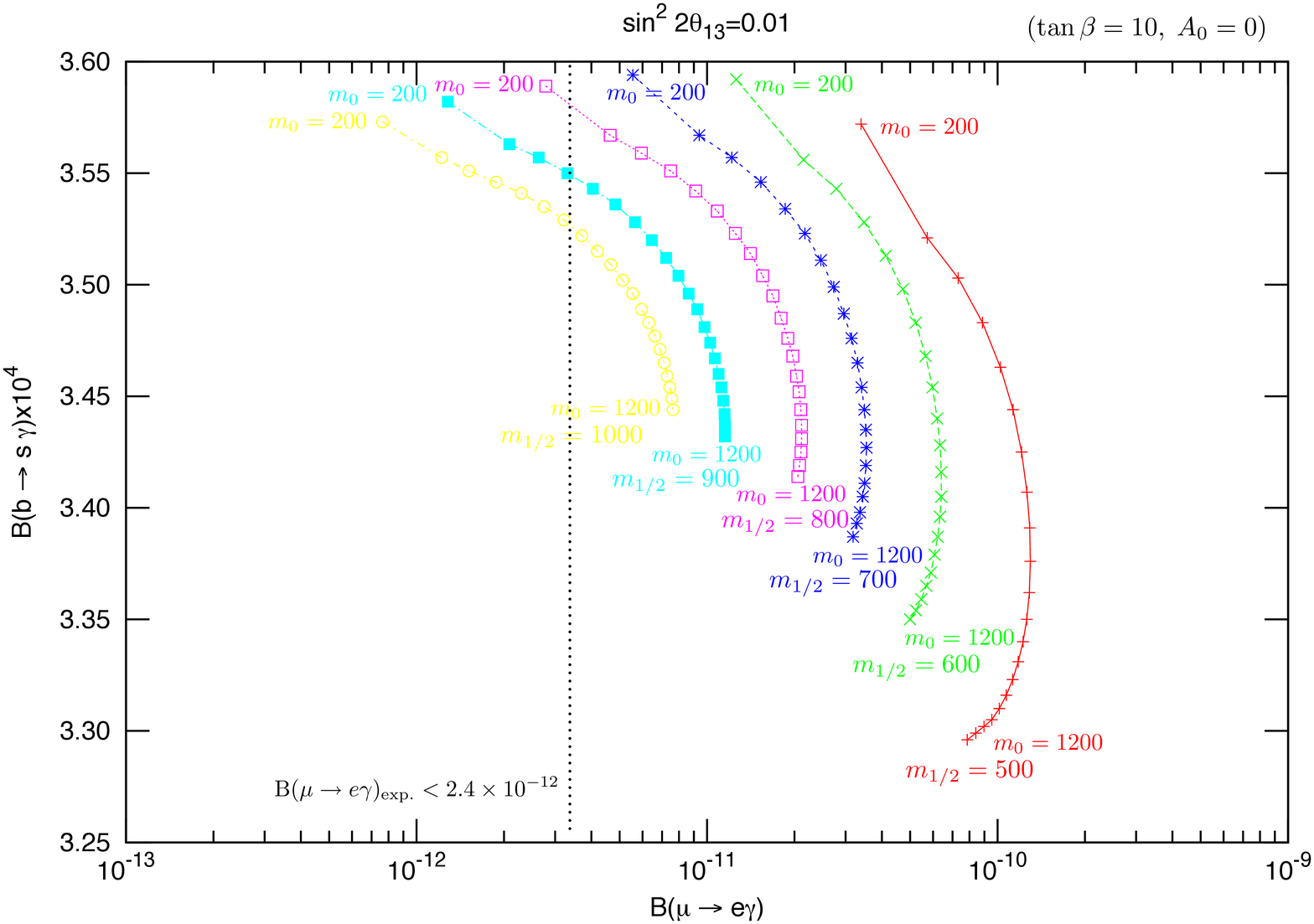}
\caption{\footnotesize Contour plot of ${\rm B}(b\to s\gamma)$ and ${\rm B}(\mu\to e\gamma)$ with $\sin^22\theta_{13}=0.01$. We take $500{\rm GeV}\leq m_{1/2}\leq 1000{\rm GeV}$. Other parameters are the same as in Fig. \ref{meg-bsg-0}.}
  \label{meg-bsg-001}
\end{minipage}
\end{figure}

Let us show results of numerical analyses in    
 $\mu\to e\gamma$, $\tau\to \mu\gamma$, and $b\to s\gamma$ processes.
Figures \ref{meg-bsg-0} and \ref{meg-bsg-001}
%, and \ref{meg-bsg-01}
show correlations between
 branching ratios of ${\rm B}(b\to s\gamma)$ and ${\rm B}(\mu\to e\gamma)$
 with $\tan\beta=10$ and $A_0=0$.
%{\color{cyan}
In Fig.\ref{meg-bsg-0},
% $m_{1/2}$ of each line is varied from 500 GeV in the right-hand line 
% to 800 GeV in the left-hand line increasing by 100 GeV unit (?).
$m_{1/2}$ is varied from 500 GeV to 800 GeV by 100 GeV.
%}
As for Fig.\ref{meg-bsg-001},
%the two lines in left-hand side are $m_{1/2}=900$ GeV and 1000 GeV.
%{\color{magenta}
$m_{1/2}$ is varid from 500 GeV to 1000 GeV by 100 GeV.
%}
$m_0$ is varied from 200 GeV to 1200 GeV by 100 GeV for each line.
%In this SUSY parameter region,
Here, the Higgs mass, 
%{\color{cyan}
calclated by {\it FeynHiggs} 
\cite{Heinemeyer:1998yj,Heinemeyer:1998np,Frank:2006yh,Degrassi:2002fi},
%}
 is varied around 118 GeV
which is not excluded by ATLAS \cite{HiggsAtlas} and CMS \cite{HiggsCMS}.
In Figs. \ref{meg-bsg-0} and \ref{meg-bsg-001},
%, and \ref{meg-bsg-001}
$\sin^22\theta_{13}$ is taken by 0 and 0.01, respectively.
We consider that spectrum of neutrinos is hierarchical,
and the $\nu_\tau$-Yukawa coupling is of ${\cal O}(1)$.
The current upper bound on ${\rm B}(\mu\to e\gamma)$ is
$2.4 \times 10^{-12}$ by the MEG experiment \cite{Adam:2011ch}.
Figure \ref{meg-bsg-001} shows 
%{\color{magenta}
that 
%}
large $\theta_{13}$ is restricted in $\mu\to e\gamma$.
In this parameter region, 
${\rm B}(b\to s\gamma)$ does not change 
%{\color{magenta}
drastically
because $m_{1/2}$-dependence is larger than 
$m_0$-dependence.
%}

\begin{figure}[htbp]
  \def\@captype{table}
  \begin{minipage}[r]{.48\textwidth}
  \includegraphics[width=80mm]{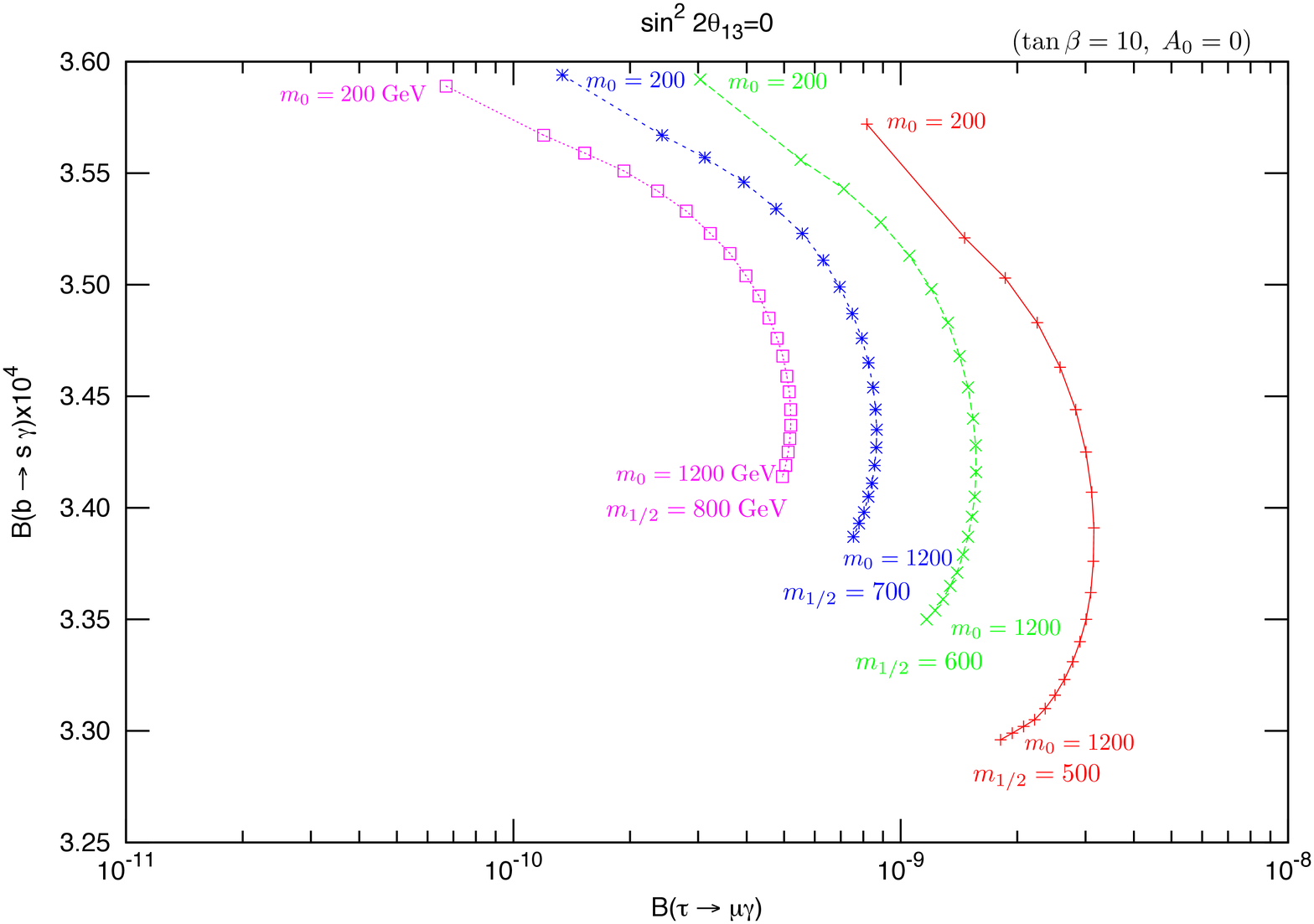}
\caption{\footnotesize Contour plot of ${\rm B}(b\to s\gamma)$ and ${\rm B}(\tau\to \mu\gamma)$ with $\sin^22\theta_{13}=0$. Here, we take $200{\rm GeV}\leq m_0\leq 1200{\rm GeV}$, $500{\rm GeV}\leq m_{1/2}\leq 800{\rm GeV}$, $A_0=0$, and $\tan\beta=10$. Experimental upper bound for ${\rm B}(\tau\to \mu\gamma)$ is $4.4\times 10^{-8}$.}
 \label{tmg-bsg-0}
\end{minipage}
  \hfill
  \begin{minipage}[l]{.48\textwidth}
  \includegraphics[width=80mm]{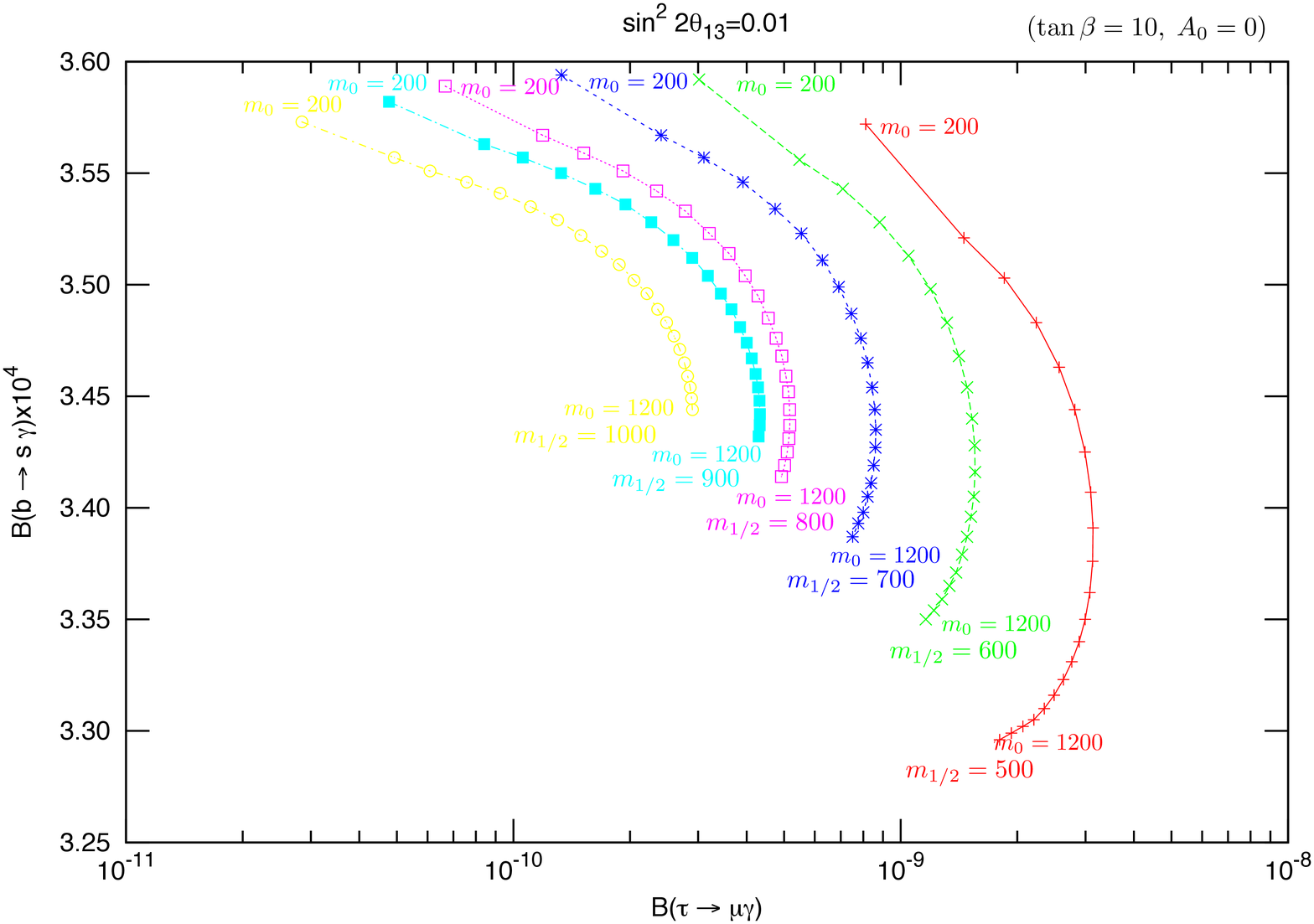}
\caption{\footnotesize Contour plot of ${\rm B}(b\to s\gamma)$ and ${\rm
   B}(\tau\to \mu\gamma)$ with $\sin^22\theta_{13}=0.01$. We take
   $500{\rm GeV}\leq m_{1/2}\leq 1000{\rm GeV}$. Other parameters are
   the same as in Fig. \ref{tmg-bsg-0}.}
  \label{tmg-bsg-001}
\end{minipage}
\end{figure}

Figures \ref{tmg-bsg-0} and \ref{tmg-bsg-001} 
show correlations between ${\rm B}(b\to s\gamma)$ and ${\rm B}(\tau\to \mu\gamma)$,
which parameters are the same as
 Figs.\ref{meg-bsg-0} and \ref{meg-bsg-001}, respectively.
${\rm B}(\tau\to \mu\gamma)$ does
 not reach the experimental 
%{\color{magenta}
upper
%}
 bound in this parameter region.
(The experimental 
%{\color{magenta}
upper 
%}
bound for ${\rm B}(\tau\to \mu\gamma)$ is $4.4\times 10^{-8}$ 
by BABAR experiment \cite{:2009tk}.) 
%{\color{cyan}
Note that a ratio of ${\rm B}(\tau\to \mu\gamma)/{\rm B}(\mu\to e\gamma)$
 depends largely on $\theta_{13}$,  
 where other neutrino oscillation parameters are fixed.
%Therefore, once $\theta_{13}$ is fixed, 
%the ratio can be predicted.
%}
When $\theta_{13}$ becomes large,
${\rm B}(\tau\to \mu\gamma)/{\rm B}(\mu\to e\gamma)$ closes to 10.
This behavior is consistent with Ref. \cite{Kitano:2002px}.

We do not consider $\tau\to e\gamma$ process 
because the experimental 
 %{\color{magenta}
upper
%}
 bound for ${\rm B}(\tau\to e\gamma)$ is 
$3.3\times 10^{-8}$ \cite{:2009tk}
which is the same order as ${\rm B}(\tau\to \mu\gamma)$.
Ratio of ${\rm B}(\tau\to e\gamma)/{\rm B}(\tau\to \mu\gamma)$
is roughly proportional to $(V_D)_{31}^2/(V_D)_{32}^2<1$.
Hence, ${\rm B}(\tau\to \mu\gamma)$ is more stringent constraint 
than ${\rm B}(\tau\to e\gamma)$.

%${\rm B}(b\to s\gamma)$ and 
%${\rm B}(\mu\to e\gamma)$ (or ${\rm B}(\tau\to \mu\gamma)$) and ${\rm B}(b\to s\gamma)$.

%\begin{figure}[htbp]
%  \def\@captype{table}
%  \begin{minipage}[c]{.48\textwidth}
%  \includegraphics[width=80mm]{meg-bsg-01}
%\caption{\footnotesize }  
%\label{meg-bsg-001}
%\end{minipage}
%  %
%  \hfill
%  %
%  \begin{minipage}[c]{.48\textwidth}
%  \includegraphics[width=80mm]{}
%\caption{\footnotesize }
% \label{}
%\end{minipage}
%\end{figure}

\begin{figure}[htbp]
\begin{minipage}[r]{.48\textwidth}
\includegraphics[width=80mm]{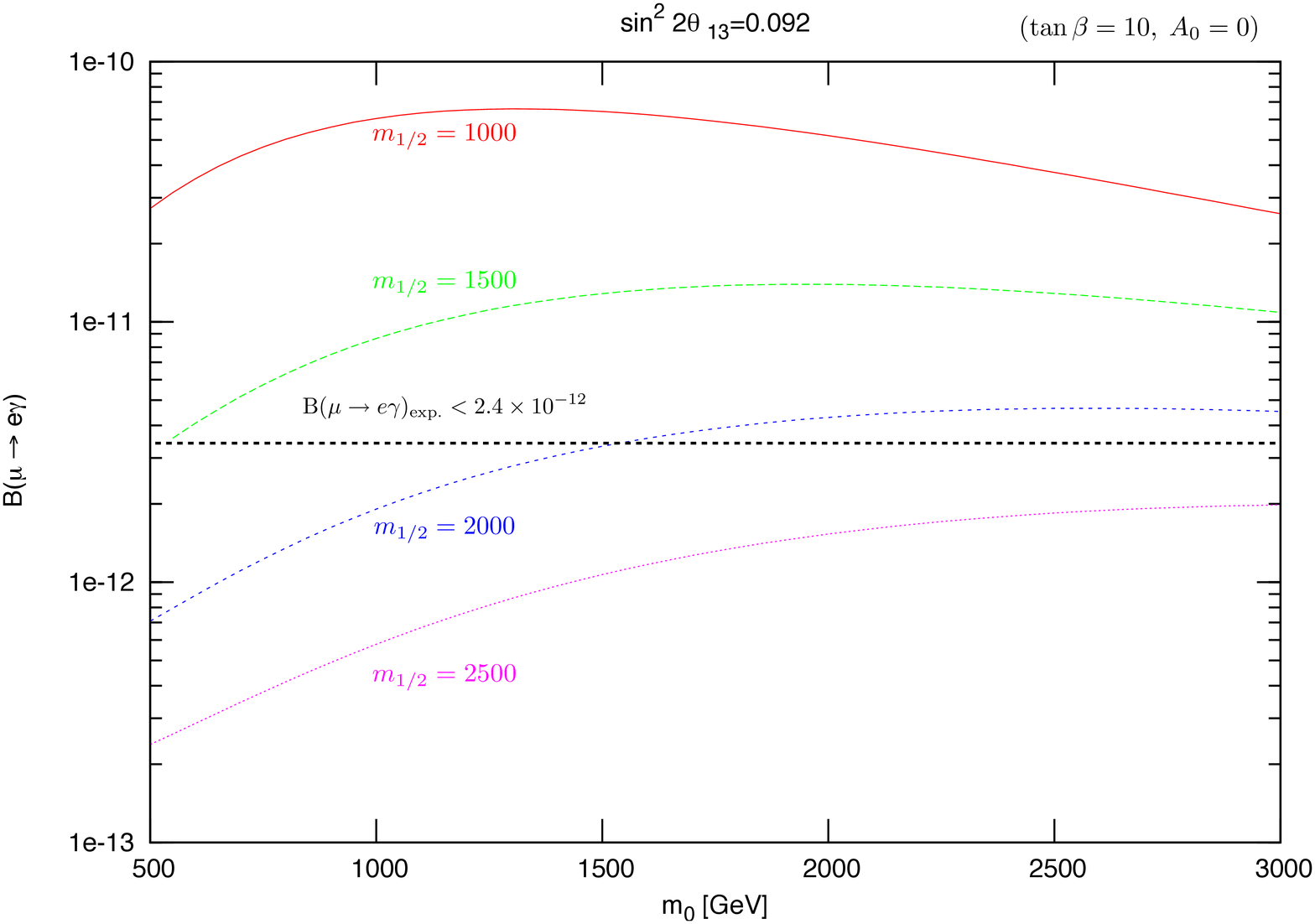}
\caption{\footnotesize ${\rm B}(\mu\to e\gamma)$ is plotted as a function of $m_0$ with $\sin ^2 2\theta_{13}=0.092$. We take $m_{1/2}=$ 1 TeV, 1.5 TeV, 2 TeV, and 2.5 TeV.}
\label{meg-db}
\end{minipage}
\end{figure}

Finally we comment on the Daya Bay experiment,
 which has measured a non-zero $\theta_{13}$ \cite{An:2012eh}.
The best-fit value is given by $\sin ^2 2\theta_{13}=0.092$,
 and such large mixing angle gives more stringent constraint in $\mu\to e\gamma$.
%{\color{cyan}
%When we take $\sin ^2 2\theta_{13}=0.092$,
% sparticles must be heavier in order not to excess 
% the experimental upper bound.
% {\color{magenta} 
Figure \ref{meg-db} shows $m_0$- and $m_{1/2}$-dependence of ${\rm B}(\mu\to e\gamma)$ with $\sin ^2 2\theta_{13}=0.092$.
We can see that $m_{1/2}$ should be larger than 2 TeV in order not to excess the experimental bound in Fig.\ref{meg-db}.
%} 
%and SUSY contribution tends to be small in flavor changing processes.
%(This constraint does not mean that SUSY $SU(5)_{H_\nu}$ model is excluded 
%in large $\theta_{13}$.)
%}
As for neutrinoless double beta decay,
it is forbidden in our setup
 because neutrinos are Dirac fermion with  
% and there are no sources inducing
 lepton number conservation.

\section{Summary}

Among three typical energy scales, 
 a neutrino mass scale, 
 a GUT scale, 
 and  
 a TeV (SUSY)-scale, 
 there is a marvelous relation of Eq.(\ref{1}). 
In this paper, 
%{\color{magenta}
we have investigated phenomenology of
a SUSY 
%{\color{magenta}
$SU(5)_{H_\nu}$ model 
%}
proposed in Ref. \cite{Haba:2011pm}.
This model %}
% we have suggested  
% a simple supersymmetric neutrinophilic 
% Higgs doublet model, which 
 realizes the relation 
 of Eq.(\ref{1})  
 dynamically as well as  
 the suitable $m_\nu$ through  
 a tiny VEV 
 of neutrinophilic Higgs.  
% without 
% additional scales other than 
% $M_{NP}$ and $M_{GUT}$.     
%Usually,
% SUSY neutrinophilic doublet models have %, 
% tiny mass scale of 
% soft $Z_2$-symmetry breaking as 
% $\rho,\rho'={\cal O}(10)$ eV.  
%This additional tiny mass scale plays a crucial role of 
% generating the tiny neutrino mass, %in a 
% usual neutrinophilic Higgs doublet model, 
% however, 
%% we did not argue 
% its origin %of it %this scale
% is just an assumption. 
%In other words, 
% the smallness of $m_\nu$ is just replaced by 
% that of $Z_2$-symmetry breaking mass parameters, 
% and this is not an essential explanation of 
% tiny $m_\nu$.   
%This is a common serious problem exists in 
% neutrinophilic Higgs doublet models in general. 
%Our model have solved this problem, where 
% two scales of
% $M_{GUT}$ and $M_{NP}$  
% naturally induce  
% the suitable magnitude of $m_\nu$  
% through the relation of Eq.(\ref{1}),   
% and does not require 
% any additional scales.  
%%A gauge coupling unification %, 
%% is also preserved automatically 
%% in our
%% setup. 
%{\color{magenta}
At first, we have discussed the gauge coupling unification
 and the proton stability. 
% by considering the role of $T,\bar{T}$ and $T_\nu,\bar{T}_\nu$ in this model.
%If we take 
%{\color{cyan}
Fascinatingly, 
 the $SU(5)_{H_\nu}$ can realize 
 not only
 accurate gauge coupling unification 
 but also
 enough proton stability 
 simultaneously, 
 which situation is hardly realized in usual four-dimensional 
 $SU(5)$ GUTs.  
%When $3.5\times 10^{14}\;{\rm GeV}\stackrel{<}{_\sim} M''\stackrel{<}{_\sim}3.6\times 10^{15}$ GeV
%and $M'>M_{GUT}$,
%}
Next, we have investigated correlations between $b\to s\gamma$ and 
 $\mu \to e \gamma$, $\tau\to \mu\gamma$.  
% {\color{cyan}
Notice that %  Remarkably,
  $B(b\to s\gamma),\;B(\mu\to e\gamma)$ and $B(\tau\to \mu\gamma)$
  are correlated directly through neutrino mixing in the $SU(5)_{H_\nu}$ model,
which is an advantage of this model over the $SU(5)_{RN}$ model.
As shown in Eq.(\ref{dmL-RN}),
 additional unknown degrees of freedom, 
 {\it parameters in $V_M$}, are needed in the latter model. 
%because neutrinos are Majorana.
%However, there are no degrees of freedom of $V_M$ in the $SU(5)_{H_\nu}$ model.
Therefore, flavor changing processes are strongly predicted 
 in the $SU(5)_{H_\nu}$ model. 
%Even if $V_M$ is an unit matrix,
% $\delta m^2_{\tilde{L}}$ and
% $\delta m^2_{\tilde{D}}$ 
%can be different between
%the $SU(5)_{H_\nu}$ model and the $SU(5)_{RN}$ model due to the $\log$ factor.
%Naturally, when all neutrino oscillation parameters are fixed,
%the branching ratios of $b\to s\gamma$, 
%$\mu \to e \gamma$, and $\tau\to \mu\gamma$
% can be predicted in the $SU(5)_{H_\mu}$ model. %(excepting SUSY breaking parameters).
% Meanwhile, 
As for the dependence of $\theta_{13}$, 
 $B(\mu\to e\gamma)$ depends largely on it, %$\theta_{13}$,
 so that 
 $B(\mu\to e\gamma)$ is strongly limited in large $\theta_{13}$. 
On the other hand, 
 we have shown that  
%where other neutrino oscillation parameters are fixed.
 $B(b\to s\gamma)$ does not depend largely on 
 $\theta_{13}$.

%Consequently,  
%and SUSY breaking parameters must be more large.
% }
%(What is conclusion here?) 

%and discussed the bounds for LFV in large $\theta_{13}$.

%%%%%%%%%%%%%%%%%%%%%%%%%%%%%%%%%%%%%%%%%%%%%%%%%%%%%%%%%%%%%%%%%%%%%%%%%%

\vspace{1cm}

{\large \bf Acknowledgments}\\

\vspace{-.2cm}
\noindent
We thank T. Horita 
 for collaboration in early stage of this research. 
This work is partially supported by Scientific Grant by Ministry of 
 Education and Science, Nos. 20540272, 20039006, and 20025004.

%%%%%%%%%%%%%%%%%%%%%%%%%%%%%%%%%%%%%%%%%%%%%%%%%%%%%%%%%%%%%%%%%%%%%%%%%

\end{document}